\documentclass[aps,preprint,amsmath,amssymb]{revtex4}
\usepackage{graphicx} % Include figure files
\usepackage{xcolor}
\usepackage{amsmath}
\usepackage{url}
\usepackage{hyperref}
\usepackage{amssymb}

\begin{document}

\title{Rocking, Rolling, and Hopping: Exploring the Multi-motion Capabilities of Rigid and Soft Ellipsoidal Actuators}
\author{Shih-Yuan Chen$^{1}$}
\author{Michelle M. Driscoll$^{1}$}
\affiliation{$^{1}$Department of Physics and Astronomy, Northwestern University}

%\thanks{MMD and BGC contributed equally to this work}

\begin{abstract}
The problem of a rigid disk rolling down a ramp is a classic problem given to students in introductory mechanics courses. In contrast, systematic studies on the rolling behavior of an ellipse have only recently emerged. Unlike a rolling disk, where the geometric center remains at a constant height from the floor, the center of a rotating ellipse changes nonlinearly due to its eccentric shape. This eccentricity introduces new modes of motion beyond rolling, including rocking and hopping. Leveraging this multi-motion behavior, we design an ellipsoidal actuator which exhibits both rolling and hopping behaviors in response to changes in the applied angular velocity. Using a simple geometric framework, we successfully capture the motion of the actuator as a force-driven rigid ellipsoid on a non-slip flat surface, and identify the critical angular velocity for the rolling-to-hopping transition. Furthermore, by adding deformability to the actuator, we unlock new functionalities, enabling soft actuators that can climb slopes and work together to collectively ascend stairs.
\end{abstract}

\maketitle

Driven actuators are the foundation of robot motion. The simplest actuator---a disk-shaped wheel---is an incredibly efficient way to enhance transport. Here we show that, by adding eccentricity, we create an new type of actuator that can not only roll, but also hop. Moreover, by adding extra nonlinearities by introducing a deformable actuator, we add even more functionalities to this simple system.
Our actuator is driven by external magnetic fields, a widely used approach in the soft matter community due to its suitability for a wide range of situations and environments \cite{el-atab_soft_2020, gariya_review_2023}, including biological settings \cite{mallick_review_nodate_2023}. Magnetic fields are especially useful for guiding soft actuators through vessel-like environments \cite{tierno_recent_2014} and for reshaping and controlling the behavior of soft robots \cite{yasa_overview_2023, he_magnetic_2023}. Typically, the motion of drop-like and magnetically responsive soft actuators is controlled by mechanically moving a magnet \cite{nguyen_deformation_2013}, applying a strong gradient field \cite{mohammadrashidi_experimental_2023}, or using switchable magnetic coil arrays \cite{mohammadrashidi_vibration_2023}. In contrast, our system applies a uniform, rotating magnetic field to the actuator, which we use as a controlled system to explore how the behavior of a rotating object changes with its shape and angular velocity.

\begin{figure}[ht]
    \centering
    \includegraphics[width = \linewidth]{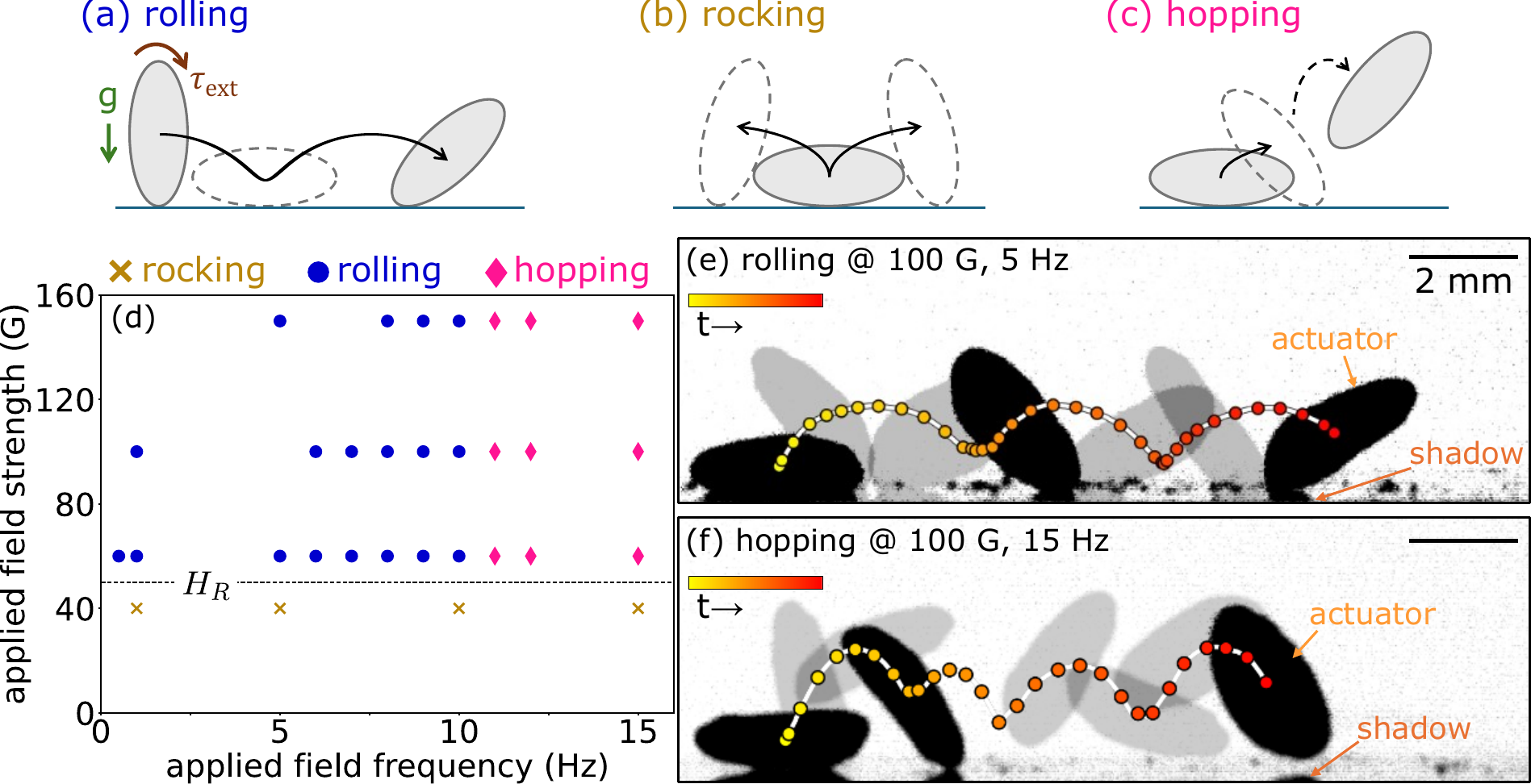}
    \caption{\textbf{Rocking, rolling and hopping of a rigid actuator.} 
    As illustrated in (a-c), a rotating ellipsoidal actuator exhibits three distinctive motions: rolling, rocking, and hopping, which depend on the applied field strength and the field frequency, as shown in the state diagram (d).
    Rocking occurs when the applied magnetic torque is below the critical threshold (corresponding to the critical applied magnetic field strength, $H_{R}$.) Above $H_{R}$, the actuator either rolls or hops, with the rolling-hopping transition depending solely on the applied frequency. In the experiments, rolling motion (e) indicates that the actuator always maintains contact with the floor, while hopping motion (f) denotes detachment from the floor, which can be visualized by separation from its shadow.
}\label{fig:rigid}
\end{figure}

Our soft actuators are ferrofluid liquid marbles \cite{aussillous_liquid_2001, tenjimbayashi_liquid_2023, bormashenko_new_2008}, drops of magnetic liquid (Ferrotec, EMG 700) coated with a shell of hydrophobic powder (Runaway Bike Store, $1-3 \ \mu$m Teflon powder). 
%To investigate the actuator's motion with external rotating fields, we compose the actuator of a ferrofluid droplet, which contains magnetic nanoparticles, and a layer of hydrophobic Teflon powder ($1-3 \ \mu$m), forming what is known as a ferrofluid marble \cite{aussillous_liquid_2001, tenjimbayashi_liquid_2023, bormashenko_new_2008}. 
We apply a DC magnetic field to horizontally stretch the ferrofluid marble (perpendicular to gravity), allowing approximately $40 \%$ of its water content to evaporate while it remains stretched. As the marble deflates, its surface crumples \cite{dandan_evaporation_2009, tosun_evaporation_2009}, causing the powder particles to jam together and form a stiff outer shell, resulting in what we refer to as a rigid actuator. 
We then place the rigid actuator on a glass slide at the center of two pairs of Helmholtz coils and apply a rotating magnetic field with the axis of rotation perpendicular to gravity. The detail of the experimental setup is provided in Methods section. The rotating magnetic field exerts a magnetic torque $\tau_{ext}$ that rotates the actuator against gravitational torque (see Fig.~\ref{fig:rigid}(a)). Our rigid actuator do not deform in the applied field (see Fig.~S2). 
Ferrofluid droplets with no shell have been employed as rolling actuators \cite{aggarwal_wobbling_2024}, but moving these drops requires overcoming the dissipation created by contact-line pinning of the drop to the surface.
%Recently, a rolling actuator made from magnetically responsive sessile droplet under a rotational magnetic field has been observed \cite{aggarwal_wobbling_2024}. 
Compared to the rolling droplet, our actuator avoids contact line hysteresis, as it does not adhere to the substrate, thus allowing it to move faster and more efficiently.

When a spherical actuator rotates, its geometric center is always at the same height, thus it can only roll (or slip). Conversely, an elliptical actuator has eccentric shape, and we find this nonlinearity allows for three distinct motions: rocking, rolling, and hopping (as illustrated in Fig.~\ref{fig:rigid}(a), and video 1-3 \cite{video}). Rocking refers to the actuator moving back and forth without any net displacement over a cycle (Fig.~\ref{fig:rigid}(b)). Rolling occurs when the actuator flips over and moves in the controlled direction (Fig.~\ref{fig:rigid}(a)). Hopping is defined as, when the actuator detaches from the floor (Fig.~\ref{fig:rigid}(c)), its vertical velocity remains positive, or $v_z > 0$.
Which of these modes of motion is observed is governed by both the applied field strength and the driving frequency (Fig.~\ref{fig:rigid}(d)). Below a critical field strength, $H_R$, the actuator rocks back and forth without net displacement, as the rotating field lacks the torque to overcome gravity.
To determine $H_R$, we apply a DC magnetic field along the $z$-axis, opposing gravity, and observe that the actuator stands upright only when the field strength exceeds $55$~G, closely matching the criteria observed in the state diagram (Fig.~\ref{fig:rigid}(d).) Notably, the field strength required to maintain the actuator in an upright position ($55$~G) is slightly higher than that needed for rolling ($50$~G). This difference arises because the lifting experiment is a torque-balanced state at equilibrium (with no angular velocity), whereas, in the rolling experiment, the field drives the actuator with angular velocity. As a result, the actuator surpasses the torque threshold due to its kinetic energy, enabling it to roll at a lower applied field strength.
%
%We then gradually increased the magnetic field until the actuator fully stood upright, identifying $H_R$. As shown in Fig.~\ref{fig:rigid}(b), $H_R$ successfully captures the transition from rocking to other motions. 

Beyond $H_R$, the actuator's motion depends solely on the applied field frequency. Below a critical frequency, $f_c$, the actuator synchronizes its major axis with the rotating magnetic field and rolls along the surface (Fig.~\ref{fig:rigid}(e)). Once exceeding $f_c$, the actuator transitions to hopping (Fig.~\ref{fig:rigid}(f)). Interestingly, the rolling speed is independent of both the actuator's mass and the applied field strength and is instead determined solely by its shape. This behavior suggests that the rolling motion is governed by simple geometric constraints.

\begin{figure}[!ht]
    \centering
    \includegraphics[width = \linewidth]{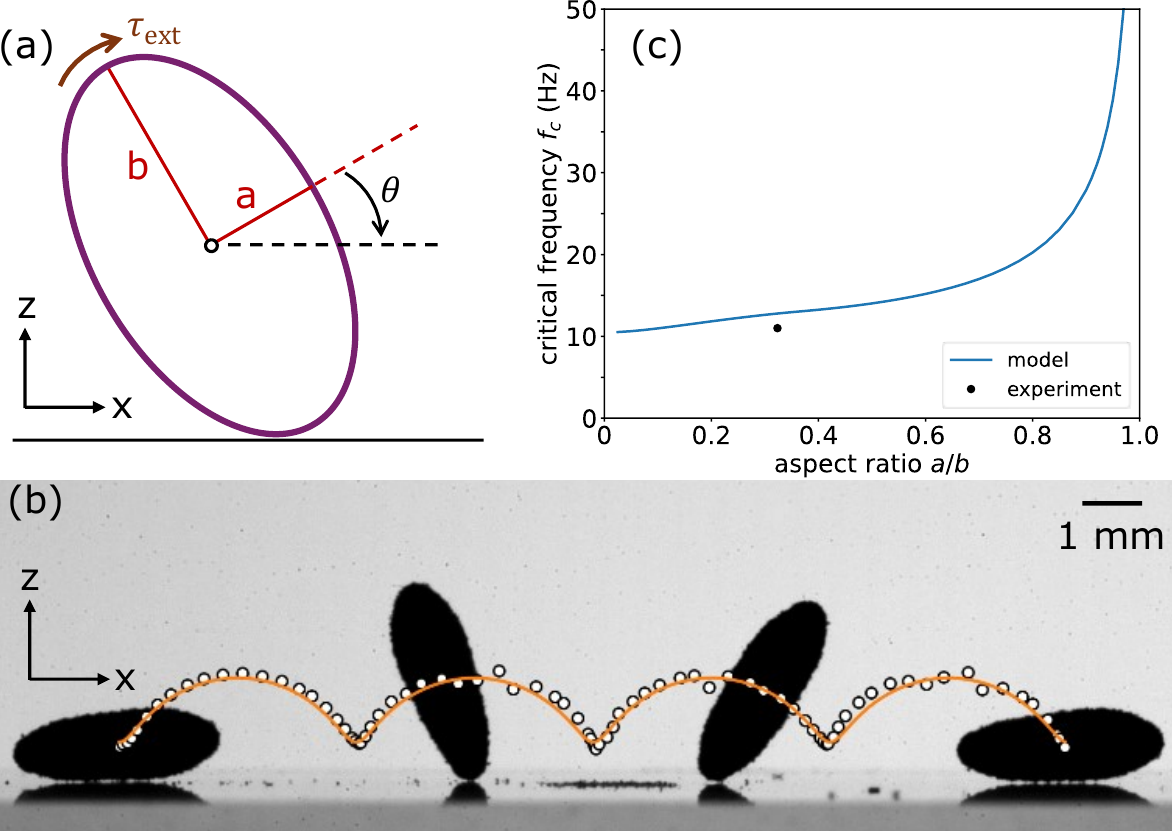}
    \caption{\textbf{Geometric model for the rolling actuator.}
    We develop a model to describe the rolling motion of a forced-rotating ellipsoid moving in one direction, as illustrated in (a), which is governed purely by geometric constraints. $a$ and $b$ are the short and long axes, and $\theta$ is the rotational angle from $a$ to the $x$ axis.
    We numerically solve the model, demonstrating excellent agreement with the experimental observation in (b). The white dots represent the geometric center of the rolling actuator, while the yellow solid line shows the model prediction, using the only fitting parameter $\eta=0.5$. 
    %The applied field frequency $f_d$ is $5$~Hz, and the shape is $a= 0.654$~mm and $b= 1.707$~mm.
    The model further predicts the critical applied frequency $f_c$ at which the actuator transitions to hopping. For a given aspect ratio with a fixed $b$, the critical frequency approaches infinity as the shape approaches to a sphere, shown in (c). Using the dimensions $a= 0.545$~mm and $b= 1.685$~mm from the phase diagram (Fig.~\ref{fig:rigid}(d)), we calculate $f_c = 12.8$~Hz, which is slightly higher than the observed critical frequency ($11$~Hz), likely due to differences between an ideal ellipsoid and the real actuator.
    %When the actuator hops, its maximum height $h$ is proportional to the driven field strength; that is, the driven power. The black solid line is guide for the eye. The error bar is smaller than the marker size.
    }
    \label{fig:theory}
\end{figure}

Therefore, we model the rigid actuator as a rolling ellipsoid.
The motion of a rolling ellipse is modeled by enforcing the geometric constraint that one point on the object must always be in contact with the floor and must not slip \cite{linden_rolling_2021, heppler_rock_2021, kajiyama_approximate_2021}. Unlike a rolling circle, the constraints are nonlinear due to the eccentric shape of the ellipse.
In our system, the actuator is driven by a rotating field at a fixed angular velocity. Therefore, we modify the motion for a rolling ellipse by introducing a driving term, $\eta$, as follows:
\begin{equation}
    \left(A(\theta, a, b)+ I\right)\ddot{\theta} + B(\theta, a, b)\dot{\theta}^2 + gC(\theta, a, b) = \eta(2\pi f_d-\dot{\theta})
    \label{equ:roll}
\end{equation}
where $I=\frac{a^2+b^2}{5}$ is the moment of inertia for an ellipsoid (without the mass term), $A$, $B$, and $C$ represent the geometric constraints that govern the rolling motion of an ellipsoid on a non-slip flat surface in one direction (see Fig.~\ref{fig:theory}(a)), and $g$ is the gravitational acceleration constant. 
$A$, $B$, and $C$ (provided in the Methods section) depend on the rotational angle $\theta$, as well as the ellipsoid's minor axis $a$ and major axis $b$, viewed from the side. 
The model ignores the third axis of the ellipsoid, as the actuator is forced to rotate independently to it. Thus, our model resembles a 2D elliptical disk, except that we must use the moment of inertia which corresponds to a 3D object.
%Since these three parameters are determined purely by geometry (the shape of the marble), we directly measured these values from the image analysis. 
The right-hand side of Eq.~\ref{equ:roll} represents the applied torque, balanced by system dissipation, where $\eta$ is the strength of the balancing torque (the only fitting parameter in Eq.~\ref{equ:roll}), and $f_d$ is the rotational frequency of the applied magnetic field from the experiments. We numerically solve this equation and compare the results with our experimental data. As illustrated in Fig.~\ref{fig:theory}(b), this simple geometric model effectively captures the rolling motion of our actuator. 
In Fig.~S3, we show that this model correctly captures both components of displacement of the actuator's geometric center in time.

To find the rolling-hopping transition, we further derive the massless normal force 
\begin{equation}
    f_{\perp}/m = \ddot{z}(\ddot{\theta}) + g    
    \label{equ:norm}
\end{equation}
where $m$ is the actuator mass, and $\ddot{z}$ is its acceleration in the $z$ direction, which can be derived from $\ddot{\theta}$ in Eq.~\ref{equ:roll}. The detail derivation of $f_{\perp}/m$ is provided in the Methods section. The condition $f_{\perp}/m = 0$ marks the transition from rolling to hopping. For a given geometry, we can compute $f_{\perp}/m$ as a function of $\theta$ and $\dot{\theta}$.
Since the actuator rotates synchronously with the applied frequency ($\dot{\theta}=2\pi f_d$), we can determine whether the actuator will hop at a given frequency. As shown in Fig.~\ref{fig:theory}(c), the critical driven frequency $f_c$ gradually increases toward infinity as the ellipsoid's shape approaches to that of a sphere, because a rolling sphere maintains a constant normal force $f_{\perp}/m = g$, which is always greater than $0$.
Using the geometric values from Fig.~\ref{fig:rigid}(b) and Eq.~\ref{equ:norm}, we calculate $f_c=12.8$~Hz, which is only slightly higher than the observed critical frequency of $11$~Hz. 
We attribute this small discrepancy to the geometrical differences between a perfect ellipsoid and the actual shape of the actuator. Additionally, our model assumes a point contact that moves smoothly along the surface of an ellipsoid. However, the real actuator has a non-smooth surface, which makes its contact bumpy and alters how the normal force supports its motion during contact.

%As the actuator hops, its trajectory becomes complex, with the height of each consecutive hop varying due to differences in the initial conditions of each hop. To simplify the analysis, we focused on the first hop, where the actuator always started from a lay-down position, and the trajectory of its geometry center had to continuously increase when it detached from the floor. We define the jump height $h$ as the maximum height of its geometric center during the hop, subtracting its major axis, or $h= y_{max} - a$. We found that $h$ is proportional to the applied magnetic field strength $H$, which differs from the rolling behavior, where the motion is independent of $H$. The cut-off of $h$ marks the transition from rocking to hopping (the green area in Fig.~\ref{fig:theory}). \textcolor{red}{(see if I can have calculate the cut-off.)}
%This can be accounted for by the fact that as the actuator deviates from the geometric model, the applied torque becomes unbalanced and accelerates the actuator until it hops. The stronger the applied torque, the higher the actuator hops. Since the applied torque follows the form $\vec{m} \cdot \vec{H}$, where $\vec{m}$ is the magnetic dipole moment of the actuator, the hopping height is therefore proportional to $H$. \textcolor{red}{(see if I can have calculate the cut-off.)}

\begin{figure}[t]
    \centering
    \includegraphics[width = \linewidth]{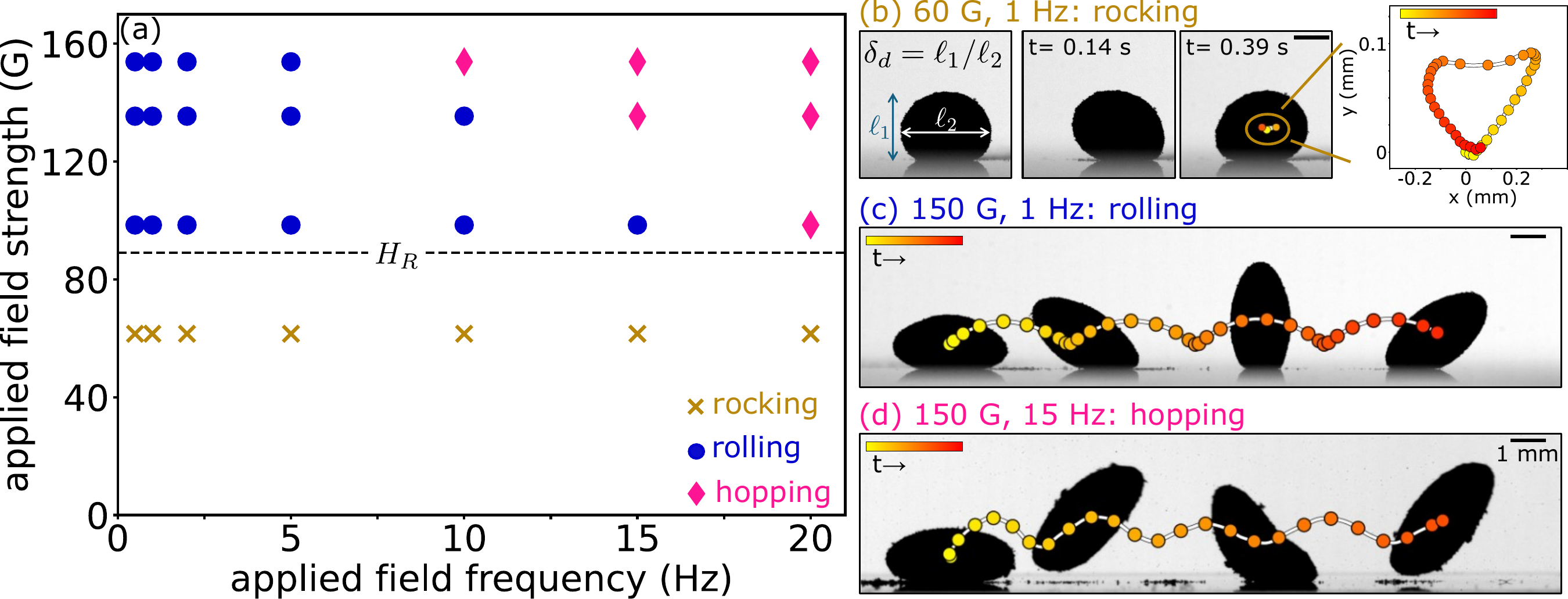}
    \caption{\textbf{Rocking, rolling, and hopping of a soft actuator.}
    The phase diagram in (a) illustrates the three distinct motion modes of a soft actuator, with corresponding visualizations in the right panels (b-d). The actuator deformation is defined as its height over its width, or $\delta_d = \ell_1/\ell2$, as shown in the first panel of (b). Rocking occurs below $H_R$, where the applied deformation is smaller than the initial deformation caused by gravity ($\delta_d<1$). As a result, the net displacement is zero, as shown in the last panel of (b). Once the applied deformation dominates, the actuator transitions to either rolling (c) or hopping (d), depending on both the applied frequency and field strength.
 }
    \label{fig:soft}
\end{figure}

%========================================================================
% soft
%========================================================================
We control the rolling-to-hopping transition by adding further nonlinearity to the system in the form of deformability.
%To control the rolling-hopping transition, we design a soft actuator. 
By reducing the rigidity of the shell, we can in situ reshape the actuator in response to the applied field, thereby controlling the transition. To maintain the actuator's deformability, we omit the evaporation step during the assembly process. 
We again observe the three modes of motion: rocking, rolling and hopping (see Fig.~\ref{fig:soft} and video 4-6 \cite{video}), similar to those exhibited by the rigid actuator. However, the critical applied field strength $H_R$ for rolling a soft actuator is higher than that of a rolling rigid actuator.
To determine the criteria for the rocking motion, we apply a DC magnetic field to the soft actuator and measure its deformation in response to the magnetic field, $\delta_d$ (its height over its width.) As shown in Fig.~S4, $\delta_d<1$ at low applied field strength because gravity dominates the shape and flattens the actuator. When $\delta_d = 1$, the applied deformation balances the initial deformation due to gravity, and the actuator becomes spherical, which initiates the rolling motion. 
We experimentally find $\delta_d = 1$ at $H_R = 88 \pm 12$~G (see Fig.~S4), which aligns with the observed critical field strength in Fig.~\ref{fig:soft}(a).
We note that $H_R$ for the soft actuator is higher than the field required for the rigid actuator to rock. This can be understood because the soft actuator deforms in response to the applied field; thus, the change in free energy due to the applied magnetic field not only rotates the magnetic dipole moment of the soft actuator but also deforms its shape. This deformation also modifies the critical frequency, $f_c$, for the rolling-hopping transition; that is, the soft actuator elongates with a higher applied magnetic field, shifting the transition to a lower frequency. 
We estimate $f_c$ for the rolling-hopping transition based on the actuator's shape at each field strength, as shown in Fig.~S5. As the actuator elongates under higher field strengths, its aspect ratio approaches $0$, leading to a decrease in $f_c$. While our model predicts $f_c$ to be within the same order of magnitude as the experiments, the observed decreasing trend deviates from the model. This discrepancy can be attributed to the continuous deformation of the soft actuator, influenced by a combination of surface energy, elasticity, mass, and the magnetic field—factors that extend beyond the scope of our simple geometric model.

\begin{figure}[!ht]
    \centering
    \includegraphics[width = \linewidth]{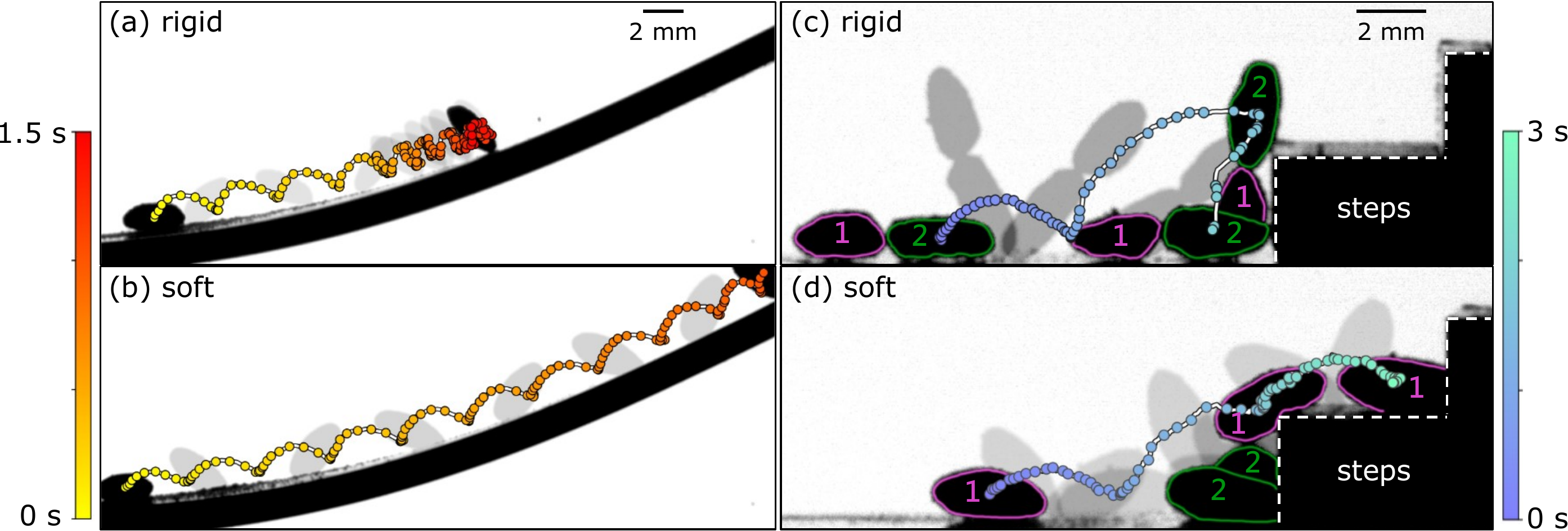}
    \caption{\textbf{New functions with deformability.} 
    Soft actuators deform to interact with complex environments, enabling them to reach locations beyond the capabilities of stiff actuators. When a stiff actuator climbs a ramp, the frictional force is insufficient to support rolling motion, causing the stiff ferromarble to slip off at a maximum height (a). In contrast, a soft actuator has a larger frictional force due to its greater contact area, allowing it to roll up the entire ramp (b). In collective motions, two connected rigid actuators attempt to climb a step but slip off (c). Meanwhile, the soft actuators pile up, pushing the upper one to reach the next step (d). The two actuators are labeled with numbers and outlined in color. The circles in all figures indicate the geometric centers of the actuators, and the colormap represents time evolution.}
    \label{fig:rigidvssoft}
\end{figure}

While soft actuators require more complex models for a quantitative understanding of their behavior, their inherent deformability introduces new functionalities and broadens their potential applications. For instance, a rigid actuator rolling uphill often slips off due to insufficient frictional force (see Fig.4(a), video 7 \cite{video}). In contrast, a soft actuator deforms and has a larger contact area with the substrate, which results in a significantly greater frictional force \cite{krim_friction_2012}. As a result, it can easily roll up a hill that a rigid actuator cannot climb (see Fig.4(b), video 8 \cite{video}). Additionally, when interacting with complex geometries, such as steps, rigid actuators tend to slide off when they make contact with the side of the step (see Fig.4(c), video 9 \cite{video}). In contrast, soft actuators adapt their shape to the step, enabling successful climbing (see Fig.4(d), video 10 \cite{video}). Moreover, collections of these soft actuators can climb over each other, effectively "boosting" one another over obstacles in their path (see video 11 \cite{video}).

In this work, we demonstrate a new type of simple actuator with multi-motion capabilities. %by changing its shape from spherical to ellipsoidal. 
We model and control its motion, and demonstrate that this ellipsoidal actuator 
can navigate through complex environments such as mazes (see video 12 \cite{video}). Notably, our model does not specify the source of the driving force, suggesting that other forces, such as rotational electric fields, could produce similar behaviors. 
Our actuator serves not only as a model for studying fundamental physics---specifically, how nonspherical objects move---but also opens avenues for investigating soft robotic control and navigation in complex environments \cite{li_soft_2022}, collective behavior in active/driven soft granular systems \cite{han_spinning_2016, narayan_long-lived_2007, kumar_flocking_2014}, and complex interactions between materials with elasticity, plasticity, and porosity \cite{draper_mapping_2019}. 
For instance, when two actuators come into contact, depending on the contact area and duration, they either bounce off each other or merge into a larger actuator. Studying this complex merging behavior reveals how to tune the actuator's speed by merging them into new shapes, as well as how liquid propagates and redistributes in porous media. Alternatively, a rapidly rotating field eventually overcomes the elastic shell of the actuator held by surface tension, splitting the actuator in half \cite{aussillous_shapes_2004}. Understanding how to control and initiate this splitting would allow us to regulate the size of the actuator and generate new actuators in situ.

\textbf{Experiments.} The actuator is made of commercial ferrofluid (Ferrotec, EMG 700, fluid density $\rho = 1290\textrm{ kg/m}^3$, viscosity $\mu = 5$~cP, saturation magnetization $M_s = 355$~G) wrapped with polytetrafluoroethylene (PTFE) Teflon powder (Runaway Bike Store, $1.6-3 \ \mu$m). For each actuator, we drop a $7.5 \ \mu$L ferrofluid droplet on a powder bed, and roll the bed until the droplet is coated with Teflon powder. For a rigid actuator, we applied a horizontal constant magnetic field ($100$~G) to stretch the actuator for $15$ minutes while letting the water content evaporate. The mass of an actuator goes down by $40 \%$ during the evaporation step. The surface of the actuator crumbles and stiffens during the process, which fixes the shape. We then use a tweezer to gently scoop the rigid actuator to a microscope glass slide and transfer the slide to the center the Helmholtz coils. For a soft actuator, we skip the evaporation process. The soft actuator is roughly $7.0 \ \mu$L in size, corresponding to \textcolor{black}{$2.5$~mm in radius when it is first deposited on the glass slide. For a rigid actuator, the length of the two short axes are between $0.5$~mm to $1.0$~mm, and the long axis is $\gtrsim  1.5$~mm.}

We drive the current from two bipolar amplifiers (Kepco BOP 50-8ML) to two orthogonal pairs of Helmholtz coils. In Fig.~S1, we illustrate the Helmholtz coils setup with a real picture. The first pair provide the magnetic field in the $z$ direction, and the second pair provide the field in the $x$ direction. The maximum magnetic field with this setup is $160$~G. 
We then use a DAQ device (Measurement Computing USB-1208HS-4AO) to control the phase between the current sent to two coils to generate a rotational magnetic field. In all experiments presented in the paper, the range of the field strength is $10-150$~G and $0.5-20$~Hz.

We use a high-speed camera (Phantom, VEO 640) to capture the actuator's motion, with recording frame rates ranging from $100$ to $2500$ fps, depending on the applied field frequency. The camera is slightly tilted ($0.8^{\circ}$) to capture the reflection of the actuators, which helps the image analysis process. ImageJ and custom Python code are used to measure the actuator's motion.

For the ramp experiments, we bend a thin Teflon slide ($1.6$~mm thick and $20$~mm wide) and fix its shape with a metal block. For the step experiments, we 3D-print the steps and fix the steps on a glass slide with double-sided tape. The height of the first step is higher than the other steps, about $3.3$~mm, due to the thickness of the double-sided tape, and the height of the second and the third ones are $3$~mm. The width of the steps is $20$~mm.

\noindent \textbf{Model.}
For a given short axis $a$ and long axis $b$ (see fig~\ref{fig:theory}), with its eccentricity $\epsilon^2= 1-a^2/b^2$, the geometric constraints $A$, $B$, and $C$ are \cite{heppler_rock_2021}
%\begin{align}
%    A &= \dfrac{a^4 \sin{\theta}^2 + b^4 \cos{\theta}^2}{a^2 \sin{\theta}^2 + b^2 \cos{\theta}^2} \\
%    B &= \dfrac{a^2 b^2 (b^2-a^2) \sin{2\theta}}{2(a^2\sin{\theta}^2+b^2\cos{\theta}^2)^2} \\
%    C &= \dfrac{(a^2-b^2)\sin{2\theta}}{2 %\sqrt{a^2\sin{\theta}^2+b^2\cos{\theta}^2}}
%\end{align}
\begin{align*}
    A &= b^2 \dfrac{1+(1-\epsilon^2)^2\tan^2{\theta}}{1+(1-\epsilon^2)\tan^2{\theta}} \\
    B &= b^2 \epsilon^2 \dfrac{(1-\epsilon^2)\tan{\theta}(1+\tan^2{\theta})}{\left(1+(1-\epsilon^2)\tan^2{\theta}\right)^2} \\
    C &= b \epsilon^2 \dfrac{\tan{\theta}}{\left(1+\tan^2{\theta}\right)\sqrt{1-\epsilon^2\sin^2{\theta}}}
\end{align*}
Note that the angle $\theta$ is defined as the angle between the minor axis $a$ to the $x$ axis. Thus, $\theta = 0$ when the ellipsoid stands up (the major axis is parallel to the $z$.) When $a=b$ (representing a circle), $B$ and $C$ go to $0$, and $A=a^2$, which agrees with the equation of motion of a rolling disk on a flat surface.
We solved Eq.~\ref{equ:roll} with Python package scipy.integrate.odeint, and set the initial condition as $\theta = \pi/2$ and $\dot{\theta}=0$. 

The massless normal force $f_{\perp}/m = \ddot{z}+g$, where $z = b\sqrt{1-\epsilon^2\sin^2{\theta}}$, can be rewritten as
\begin{equation}
    f_{\perp}/m = \dfrac{\partial^2 z}{\partial \theta^2}\dot{\theta}^2 + \dfrac{\partial z}{\partial \theta}\ddot{\theta} + g
    \label{equ:normal}
\end{equation}
Replacing $\ddot{\theta}$ using Eq.~\ref{equ:roll}, we have
%\iffalse
%Replacing $\ddot{\theta}$ using Eq.~\ref{equ:roll}, we have
\begin{equation*}
    f_{\perp}/m = \frac{\alpha + \beta}{\Delta}  
\end{equation*}
where
\begin{align*}
    \alpha &= \epsilon^2 \dot{\theta}^2 \dfrac{2 \sqrt{2} b (4 (7 - 7 \epsilon^2 + 3 \epsilon^4) \cos{2 \theta} - 3 \epsilon^2 (-2 + \epsilon^2) (3 + \cos{4\theta}))}{\sqrt{2 - \epsilon^2 + \epsilon^2 \cos{2 \theta}} } \\
    \beta &= - 56g  -g \epsilon^2(56  - 19 \epsilon^2 + 24(-2 + \epsilon^2) \cos{2 \theta} - 5 \epsilon^2 \cos{4 \theta}) \\
    \Delta &= -56 + 8 \epsilon^2 \left(7 - 3 \epsilon^2 + 3 (-2 + \epsilon^2) \cos{2 \theta} \right)
\end{align*}
When $a=b$, $f_{\perp}/m=g$ and is always larger than $0$.
When $a \neq b$, we numerically solve Eq.~\ref{equ:normal} to identify the minimum value of $\dot{\theta}=f_c$ that gives $f_{\perp}/m=0$ for different aspect ratio ($a/b$), as shown with the blue solid line in Fig.~\ref{fig:theory}. 

\bibliographystyle{unsrt} 
\bibliography{SYC_ref} 

\section*{Acknowledgement}
This work was partially supported by the University of Chicago Materials Research Science and Engineering Center, which is funded by National Science Foundation under award number DMR-2011854.
We thank Addison Benz, Natalya Guiden, Chaeun Park, and Adriana Castelan for their help in building the experimental setup and perfomring exploratory experiments. We are grateful for the insightful ideas from Abigail Plummer and Hector Manuel Lopez Rios.

\end{document}

\maketitle

\section*{Supporting videos}
\begin{itemize}
    \item \textbf{Video 1: Rocking of a rigid actuator.} An ellipsoidal rigid actuator rocks back and forth on a flat surface. The applied magnetic field is $40$~G at $1$~Hz. The experiment is recorded at $500$~fps.
    \item \textbf{Video 2: Rolling of a rigid actuator.} An ellipsoidal rigid actuator rolls along a flat surface. The applied magnetic field is $10$~G at $0.5$~Hz. The experiment is recorded at $100$~fps.
    \item \textbf{Video 3: Hopping of a rigid actuator.} An ellipsoidal rigid actuator hops on a flat surface. The applied magnetic field is $150$~G at $15$~Hz. The experiment is recorded at $500$~fps.
    \item \textbf{Video 4: Rocking of a soft actuator.} An ellipsoidal soft actuator rocks back and forth on a flat surface. The applied magnetic field is $40$~G at $1$~Hz. The experiment is recorded at $1500$~fps.
    \item \textbf{Video 5: Rolling of a soft actuator.} An ellipsoidal soft actuator rolls along a flat surface The applied magnetic field is $100$~G at $10$~Hz. The experiment is recorded at $500$~fps.
    \item \textbf{Video 6: Hopping of a soft actuator.} An ellipsoidal soft actuator hops on a flat surface The applied magnetic field is $150$~G at $15$~Hz. The experiment is recorded at $2500$~fps.
    \item \textbf{Video 7: Stiff actuator climbing up a ramp.} A stiff actuator climbs and slips off the ramp. The applied magnetic field is $100$~G at $5$~Hz. The experiment is recorded at $500$~fps.
    \item \textbf{Video 8: Soft actuator climbing up a ramp.} A soft actuator climbs through the whole ramp. The applied magnetic field is $100$~G at $5$~Hz. The experiment is recorded at $500$~fps.
    \item \textbf{Video 9: Two stiff actuators climbing steps.} Two stiff actuators self-chain together. They approach but slip off the step when contacting the step. The applied magnetic field is $150$~G at $0.5$~Hz. The experiment is recorded at $250$~fps.
    \item \textbf{Video 10: Two soft actuators climbing steps.} Two soft actuators self-chain together. They approach the step, and the first actuator pushes the second actuator to the upper step. The applied magnetic field is $150$~G at $0.5$~Hz. The experiment is recorded at $250$~fps.
    \item \textbf{Video 11: Collective behavior} Actuators pile up together to climb steps. The applied magnetic field is $100$~G at $0.5$~Hz. The experiment is recorded at $150$~fps.
    \item \textbf{Video 12: Actuator moving in a maze.} We control the direction of the rotational magnetic field with a joystick, and thus control the direction of the actuator motion. The maze is 3D-printed and fixed on a glass plate. The applied magnetic field is $120$~G at $0.5$~Hz. The experiment is recorded at $4$~fps.
\end{itemize}

\newpage
\section{Experimental setup}
\begin{figure}[ht]
    \centering
    \includegraphics[width = \linewidth]{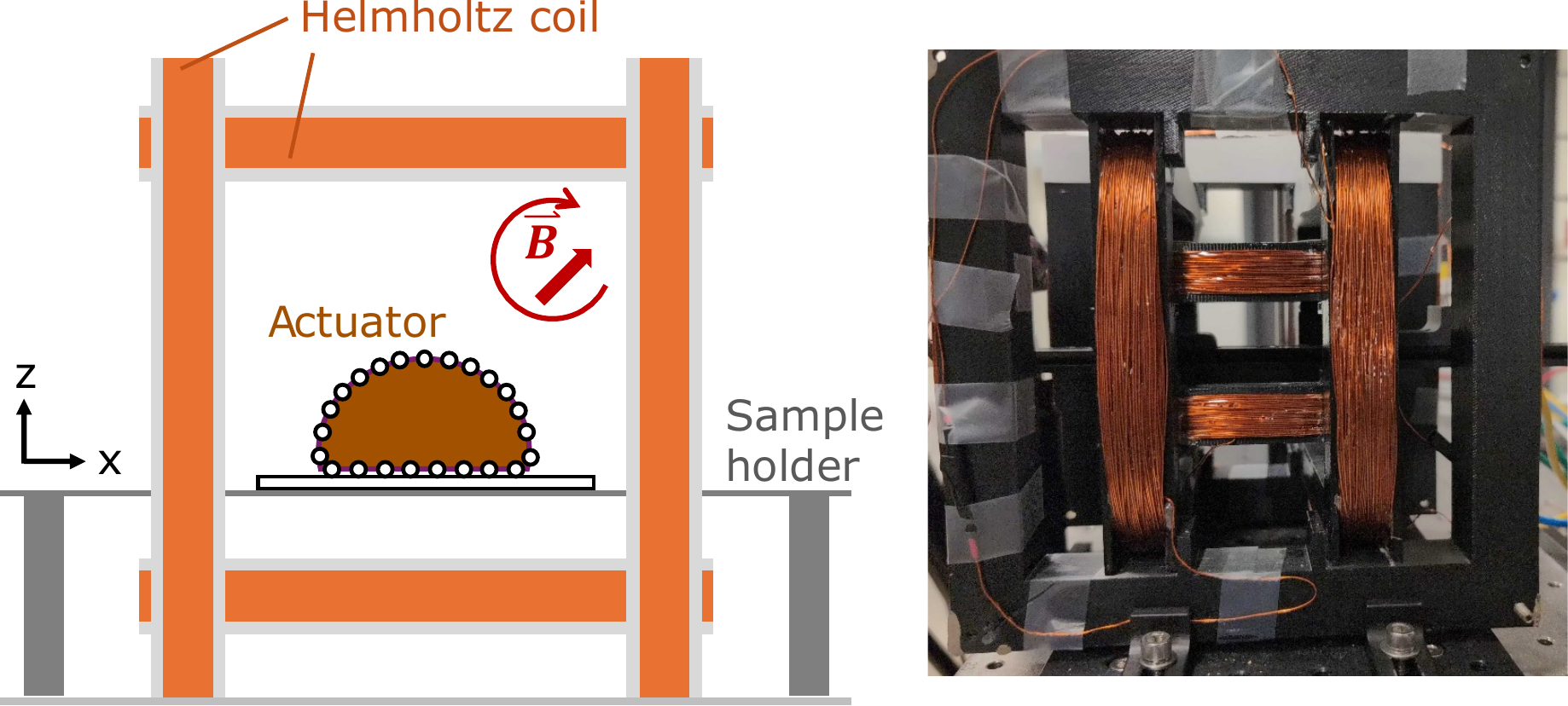}
    \caption{\textbf{Rotational magnetic fields setup.}
    The two pairs of Helmholtz coils are illustrated at the left panel, and a photograph at the right panel.
    }
    \label{figS:experiment}
\end{figure}

\newpage
\section{Rigid actuator and soft actuator under a DC magnetic field.}
\begin{figure}[ht]
    \centering
    \includegraphics[width = \linewidth]{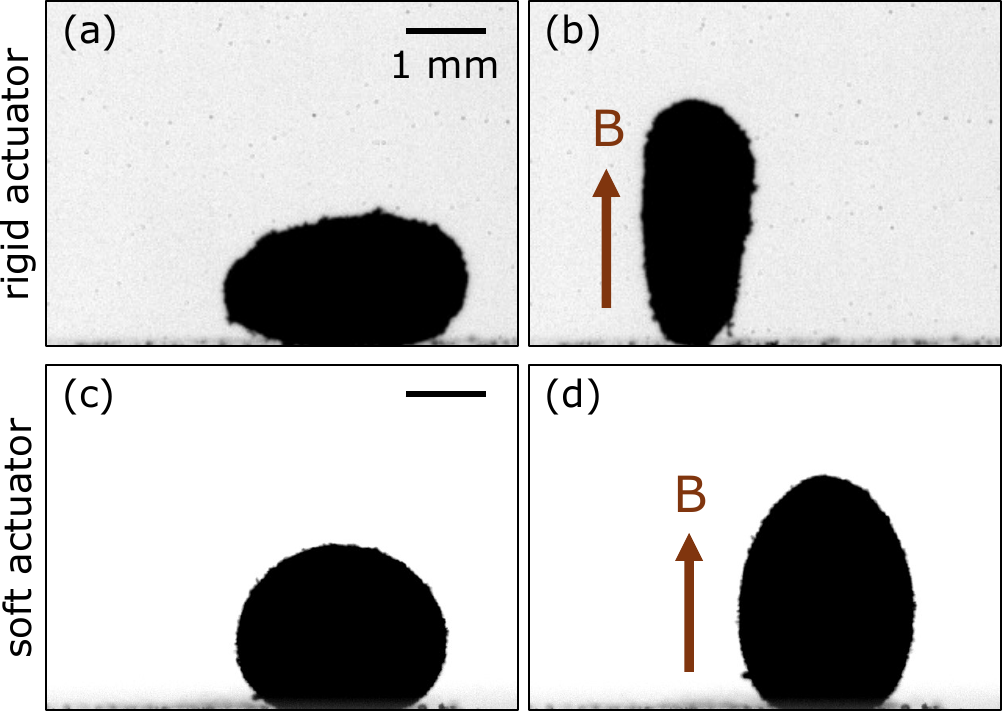}
    \caption{\textbf{Comparison between rigid and soft actuator.}
    A rigid actuator rotates and lines up with the applied DC magnetic field direction (a-b) while a soft actuator stretches itself along the applied field direction (c-d). The applied field strength is $150$~G for the rigid actuator and $135$~G for the soft actuator. 
    }
    \label{figS:stiff2soft}
\end{figure}

When we apply a DC magnetic field in the $z$ direction (against gravity), a rigid actuator rotates and lines up with the field direction, and a soft actuator elongates itself along the field direction.  

\newpage
\section{Geometric center of a rolling actuator}
\begin{figure}[ht]
    \centering
    \includegraphics[width = \linewidth]{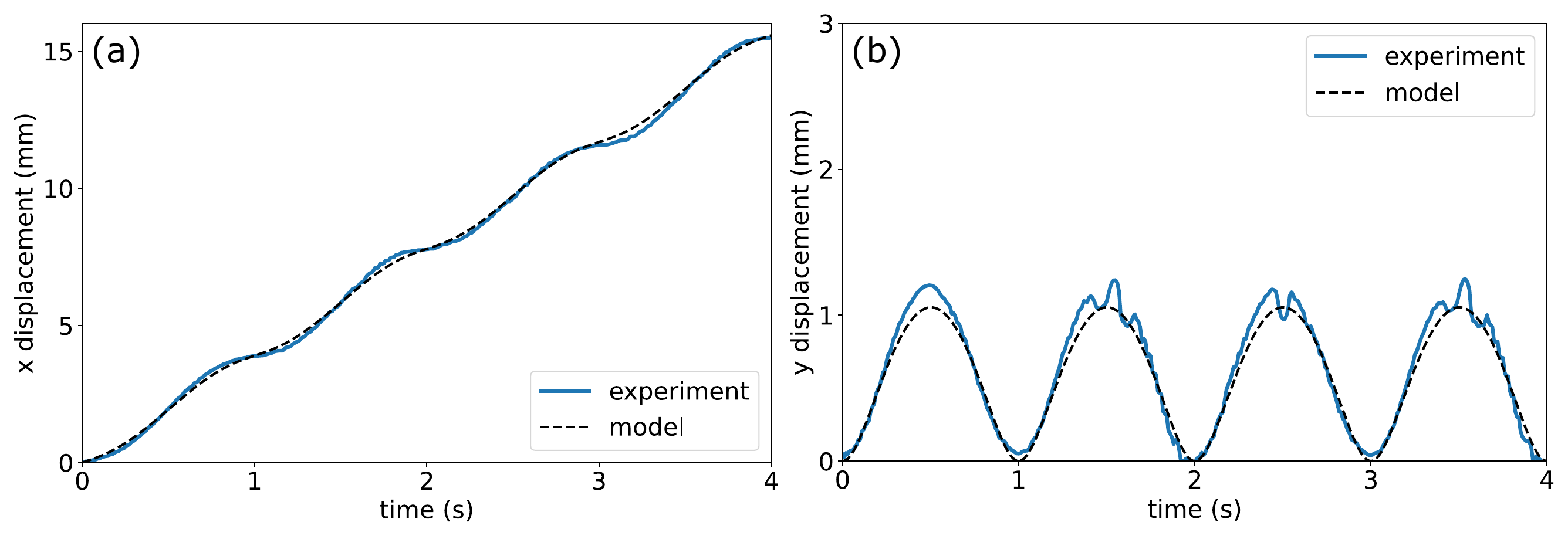}
    \caption{\textbf{Geometric model captures the rolling actuator motion.}
    The geometric model captures the center of the rolling actuator over time. The data come from fig.2.}
    \label{figS:theory}
\end{figure}

%The geometric model captures the geometric center of the rolling actuator over time. The data come from fig.2.

\newpage
\section{Deformation due to the applied field against gravity}
\begin{figure}[ht]
    \centering
    \includegraphics[width = \linewidth]{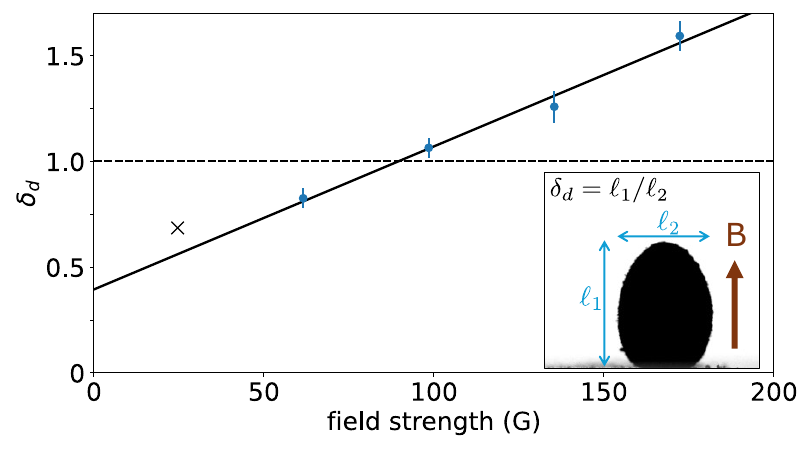}
    \caption{\textbf{Magnetic field strength threshold to dominate the actuator's shape and trigger rolling motion.} 
    The aspect ratio of a soft actuator, $\delta_d$, is initially smaller than $1$ because the gravity dominates its shape. As the applied magnetic field increases, the actuator elongates along the field direction ($\delta_d>1$). The black dashed line ($\delta_d=1$) marks the threshold of the required field strength to make the actuator spherical and roll.
}
    \label{figS:strain}
\end{figure}

The soft actuator deforms along the direction of the applied field. For the soft actuator to roll, the applied deformation due to the magnetic field must be greater than its initial deformation caused by gravity. To capture the magnetic field threshold, we calculate the actuator's aspect ratio, $\delta_d$, as its height over its width, or $\delta_d = \ell_1/\ell_2$. When $\delta_g = 1$, the actuator resembles a sphere, and it starts rolling. We fit the data with a line and find its intersection with $1$, which occurs at a field strength of $88 \pm 12$~G.

\newpage
\section{Critical frequency $f_c$ for the rolling-hopping transition of a soft actuator}
\begin{figure}[ht]
    \centering
    \includegraphics[width = \linewidth]{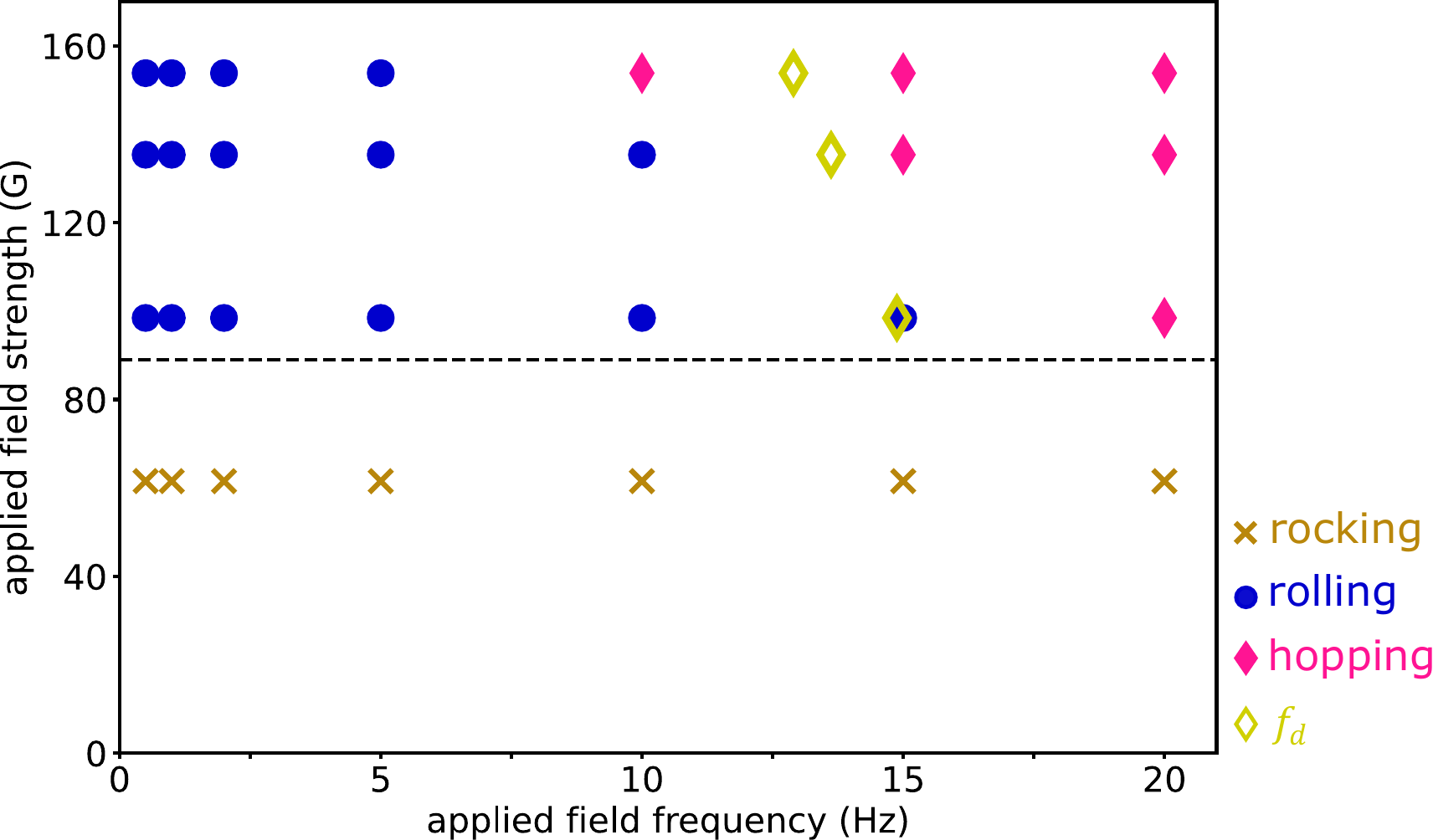}
    \caption{\textbf{Transition predicted by geometric model}
    The critical frequency for the rolling-hopping transition predicted by the geometric model is off compared to the observed data, as the shape of the soft actuator is deformed by other factors beyond the geometric model.
}
    \label{figS:fd}
\end{figure}

We fit an ellipse to capture the shape of the soft actuator at a given magnetic field, and use the major and minor axes to calculate the critical frequency, $f_c$, for the rolling-hopping transition. We believe that the offset between the predicted and observed $f_c$ is due to the fact that the actuator deforms continuously as it moves along the floor, which is not captured by the point-contact model.